\documentstyle[preprint,aps]{revtex}
\def\beq{\begin{equation}}  
\def\eeq{\end{equation}}
\tighten
\begin{document}

\title
{\bf Pairing and continuum effects in nuclei close to the drip line}

\author
{M. Grasso $^1$, N. Sandulescu $^2$, Nguyen Van Giai $^1$, R. J. Liotta
$^3$}

\address
{$^1$ Institut de Physique Nucl\'eaire, IN2P3-CNRS, 
Universit\'e Paris-Sud, 91406 Orsay Cedex, France\\
$^2$ Institute for Physics and Nuclear Engineering, 
 P.O.Box MG-6, 76900 Bucharest, Romania\\
$^3$ Royal Institute of Technology, Frescativ.24, S-10405 Stockholm, Sweden}

\maketitle

\begin{abstract}

The Hartree-Fock-Bogoliubov (HFB) equations in coordinate
representation are solved
exactly, i.e., with correct asymptotic boundary conditions 
for the continuous spectrum. The calculations are performed with
effective Skyrme interactions.
The exact HFB solutions are compared
with HFB calculations based on box boundary conditions
and with resonant continuum Hartree-Fock-BCS (HF-BCS)
results. The comparison is done
for the neutron-rich Ni isotopes. It is shown that close to the 
drip line the amount of pairing correlations depends on how 
the continuum coupling is treated. On the other hand, the
resonant continuum HF-BCS 
results are generally close to those 
of HFB even in neutron-rich nuclei.

\end{abstract}

\section{Introduction}

The physics of exotic nuclei close to the drip lines has triggered 
a new interest for the study of pairing correlations in finite systems.
A peculiarity  of pairing correlations in weakly bound  nuclei is their
sensitivity to the effects of unbound single-particle states.
 
 The pairing correlations in the presence of continuum coupling have been
 treated both in HFB
 \cite{belyaev,bulgac,doflo,dona,tehe,fay,stona,stori}
 and HF-BCS \cite{saliwy,sagili,kruppa,sacili}
 approximations.
 In the HFB approximation the continuum is generally included in 
 spherical systems by solving
 the HFB equations in coordinate representation. The calculations are
 done either in the complex energy plane by using Green function
 techniques
 \cite{belyaev,fay}, or on the real energy axis \cite{doflo,dona}. In the
 latter case the HFB equations are usually solved by imposing box boundary
 conditions, i.e., the HFB wave functions are assumed to vanish
 beyond some distance which is chosen to be typically a few times the
 nuclear radius.

 The effect of the resonant continuum upon pairing correlations was 
 also studied
 in the framework of the BCS approximation, both for zero
 \cite{saliwy,sagili,kruppa} and finite temperature \cite{sacili}.

For deformed systems working in coordinate representation is much more 
difficult \cite{tehe}. In most of the deformed HFB calculations the
continuum is discretized by expanding the HFB wave functions on a 
single-particle basis. Usually a harmonic oscillator basis is taken
and one can improve the description of physical quantities at large
distances like density tails by performing a local scaling transformation
\cite{stona,stori}.
   
 The aim of this article is to show how the coordinate space
 HFB equations can be actually solved in the case of spherical
 symmetry and Skyrme type forces
 by treating the continuum exactly, i.e., with correct boundary
 conditions,
 and to analyse to which extent different approximations,
 namely box HFB and resonant continuum HF-BCS calculations, 
 compare with the
 exact solutions. 
 In this paper we will treat the continuum exactly only for the neutrons of
 neutron-rich systems for which one expects that the continuum plays an
 important role close to the drip line. For the protons we will treat the
 continuum by a box discretization.
 It will be shown that nuclear properties related
to pairing correlations are correctly predicted by discretized continuum
methods away from drip line but they deviate appreciably from exact
continuum results when one approaches the drip line.   

 The paper is organized as follows. In Section II we give a brief reminder
 of the HFB equations in coordinate representation. In Section III
 we present the procedure we have used for calculating the continuum HFB
 solutions and we discuss, in a schematic model, how the quasiparticle
 resonant states are identified.
 In Section IV we present the continuum HFB calculations for Ni
 isotopes in comparison with box HFB and resonant HF-BCS calculations.
 Conclusions are drawn in Section V.

\section{HFB equations in coordinate representation}

The HFB approximation in coordinate representation has been discussed
quite extensively in the literature \cite{bulgac,doflo,dona} and therefore
we recall here only the basic equations.

The HFB equations in coordinate representation read \cite{doflo}:

\beq
\begin{array}{c}
\int d^{3} {\mathbf{r'}} \sum_{\sigma '} \left( \begin{array}{cc} 
h( {\mathbf{r}} \sigma , {\mathbf{r'}} \sigma' ) & \tilde{h} (
{\mathbf{r}} \sigma , {\mathbf{r'}} \sigma' ) \\ 
\tilde{h} (
{\mathbf{r}} \sigma , {\mathbf{r'}} \sigma' ) & 
-h( {\mathbf{r}} \sigma , {\mathbf{r'}} \sigma' ) \end{array} \right)
\left( \begin{array}{c} \Phi_1 (E, {\mathbf{r'}} \sigma ') \\
 \Phi_2 (E, {\mathbf{r'}} \sigma ') \end{array} \right) = \\

\left( \begin{array}{cc} E+\lambda & 0 \\ 0 & E-\lambda \end{array} \right) 
\left( \begin{array}{c} \Phi_1 (E, {\mathbf{r}} \sigma) \\
 \Phi_2 (E, {\mathbf{r}} \sigma) \end{array} \right) ~,
\end{array}
\label{1}
\eeq
where $\lambda$ is the chemical potential, 
$h$ and $\tilde{h}$ are the mean field and the pairing field, 
and $(\Phi_i)$ represents the two-component HFB quasi-particle wave function 
of energy $E$. The mean field operator $h$ is a sum of
the kinetic energy $T$ 
and the mean field potential $\Gamma$,
\beq
h({\mathbf{r}} \sigma , {\mathbf{r'}} \sigma ' )= T( 
{\mathbf{r}} , {\mathbf{r'}}) \delta_{\sigma  \sigma '} +
\Gamma ({\mathbf{r}} \sigma, {\mathbf{r'}} \sigma ') ~. 
\label{2}
\eeq

The mean field potential $\Gamma$ is expressed in terms of the
particle-hole 
two-body interaction $V$ and the particle density $\rho$ in the following way:
\beq
\Gamma({\mathbf{r}} \sigma, {\mathbf{r '}} \sigma ') =
\int d^3  {\mathbf{r_1}} d^3 {\mathbf{r _2}} \sum_{\sigma_1 \sigma  _2}
V({\mathbf{r}} \sigma , {\mathbf{r_1}} \sigma _1 ; 
{\mathbf{r'}} \sigma ' , {\mathbf{r _2}} \sigma  _2 ) 
\rho({\mathbf{r _2}} \sigma  _2 , {\mathbf{r _1}} \sigma _1 ) ~.
\label{3}
\eeq

Similarly the pairing field $\tilde{h}$ is expressed in terms of 
the pairing interaction $V_{pair}$ and the pairing density $\tilde{\rho}$:
\beq
\tilde{h}({\mathbf{r}} \sigma, {\mathbf{r'}} \sigma ') =
\int d^3 {\mathbf{r_1}} d^3 {\mathbf{r _2}} \sum_{\sigma _1 \sigma  _2}
2 \sigma ' \sigma ' _2 V_{pair}({\mathbf{r}} \sigma , {\mathbf{r'}} -
\sigma ' ; 
{\mathbf{r _1}} \sigma  _1 , {\mathbf{r _2}} -\sigma  _2 ) 
\tilde{\rho}({\mathbf{r _1}} \sigma  _1 , {\mathbf{r _2}} \sigma _2 ) ~.
\label{4}
\eeq

The particle and pairing densities $\rho$ and $\tilde{\rho}$ are defined 
by the following expressions:     

\beq
\rho( {\mathbf{r}} \sigma , {\mathbf{r'}} \sigma ') 
\equiv \sum_{0 < E_n < -\lambda} \Phi_2 (E_n , {\mathbf{r}} \sigma)
\Phi^* _2 (E_n , {\mathbf{r'}} \sigma ') +
\int_{-\lambda}^{E_{cut-off}} dE   \Phi_2 (E , {\mathbf{r}} \sigma)
\Phi^* _2 (E , {\mathbf{r'}} \sigma ') ~,
\label{5}
\eeq

\beq
\tilde{\rho}( {\mathbf{r}} \sigma , {\mathbf{r'}} \sigma ') \equiv 
\sum_{0 < E_n < -\lambda} \Phi_2 (E_n , {\mathbf{r}} \sigma)
\Phi^* _1 (E_n , {\mathbf{r'}} \sigma ') +
\int_{-\lambda}^{E_{cut-off}} dE   \Phi_2 (E , {\mathbf{r}} \sigma)
\Phi^* _1 (E , {\mathbf{r'}} \sigma ') ~,
\label{6}
\eeq
where the sums are over the discrete quasiparticle states with energies
$|E| < -\lambda$, and the integrals are over the continuous 
quasiparticle states with energies $|E| > -\lambda$. The HFB solutions
have the
following symmetry with respect to E:
\beq
\begin{array}{c}
\Phi_1 (-E, {\mathbf{r}} \sigma) = \Phi_2 (E, {\mathbf{r}} \sigma)
\\
\Phi_2 (-E, {\mathbf{r}} \sigma) = -\Phi_1 (E, {\mathbf{r}} \sigma)
\end{array}
\label{7}
\eeq
As it appears clearly from Eqs.(\ref{5}) and (\ref{6}) 
we choose to work with the positive energies.  

The particle-hole and pairing interactions in Eqs.(\ref{3}) 
and (\ref{4}) are chosen as density-dependent contact interactions, 
so that the integro-differential HFB equations reduce to  
coupled differential equations. The zero-range character of the 
pairing interaction 
is the reason why one has to adopt an energy cut-off
as seen in Eqs.(\ref{5}) and (\ref{6}). 

In this paper we consider systems with spherical symmetry. In this case 
the wave functions are readily decomposed into their radial and
spin-angular parts\cite{doflo}: \beq
\Phi _i (E, {\mathbf{r}} \sigma ) = 
u_i (Elj,r) { 1 \over r} y^{m_j}_{lj} (\hat{r}, \sigma) ~, \; i=1,2 ~,
\label{9}
\eeq
where:
\beq
 y^{m}_{lj} (\hat{r}, \sigma) \equiv Y_{lm_l}(\Theta, \Phi) \chi_{1/2}
(m_{\sigma}) (lm_l { 1 \over 2} m_{\sigma} | j m ) ~.
\label{10}
\eeq
In the following we use for the upper and lower components of the radial
wave
functions the standard notation $u_{lj}(E,r)$ and $v_{lj}(E,r)$.

 As we have already mentioned, the HFB equations are usually solved
by imposing to the radial wave functions the condition that they
vanish beyond a given distance R (box radius). In this case the
continuous spectrum is replaced by a set of discrete energies, whose
density depends on the box radius. In what follows we discuss how the
HFB equations can be solved by keeping the correct asymptotic conditions  
for the neutron wave functions.

\section { The treatment of quasi-particle continuum}

\subsection{Asymptotic behaviours}

The asymptotic behaviour of the HFB wave function is determined
by the physical condition that, at large distances the nuclear mean
field $\Gamma(r)$ and the pairing field $\Delta(r)$ vanish.
This condition requires an effective interaction of finite range and
finite-range nonlocality. Outside the range of mean fields the equations
for $\Phi _i (E, {\mathbf{r}} \sigma )$ are decoupled and one can 
readily find the asymptotic behaviour of the physical solutions
at infinity \cite{bulgac}.
Thus, for a negative chemical potential $\lambda$, i.e., for a bound
system, there are two well separated regions in the quasiparticle
spectrum.

Between 0 and $-\lambda$ the quasiparticle spectrum is discrete
and both upper and lower components of the radial HFB wave function
decay exponentially at infinity. For neutrons this implies that those
components have the form,
\begin{eqnarray}
u_{lj}(E,r) & = & A h_l^{(+)}(i\alpha_1r)~, \nonumber \\
v_{lj}(E,r) & = & B h_l^{(+)}(i\beta_1r)~,
\label{eq2}
\end{eqnarray}
where $h_l^{(+)}$ are spherical Haenkel functions,
$\alpha_1^2=-\frac{2m}{\hbar^2}(\lambda+E)$ and
$\beta_1^2=-\frac{2m}{\hbar^2}(\lambda-E)$.
These solutions correspond to the bound quasiparticle spectrum.
In this case, the solutions are normalised to unity.
 
For $E > - \lambda$ the spectrum is continuous and the solutions are:
\begin{eqnarray}
u_{lj}(E,r) & = & C [cos(\delta_{lj}) j_l(\alpha_1r)-sin(\delta_{lj})
n_l(\alpha_1r)]~, \nonumber \\
v_{lj}(E,r) & = & D_1 h_l^{(+)}(i\beta_1r)~,
\label{eq3}
\end{eqnarray}
where $j_l$ and $n_l$ are spherical Bessel and 
Neumann functions repectively and $\delta_{lj}$
is the phase shift corresponding to the angular momentum $(lj)$.
One can see that the upper component of the HFB wave function has the
standard form of a scattering state while the lower component is 
always exponentially decaying at infinity.

 The asymptotic form of the wave function should be matched with
 the inner radial wave function, which for $r \rightarrow 0$ 
 can be written as follows:

\beq
\left(
\begin{array}{c}
u_{lj}(E, r) \\ v_{lj}(E, r) \end{array} \right) = 
D_2 \left(
\begin{array}{c} r^{l+1} \\ 0 \end{array} \right) +
D_3 \left(
\begin{array}{c}  0 \\ r^{l+1}  \end{array} \right) ~,
\label{14}
\eeq

The HFB wave function is normalized to the Dirac $\delta$-function of
energy. This condition fixes the constant $C$ to the value:
\beq
C = \sqrt{ { 1 \over \pi} { 2m \over {\hbar^2 \alpha_1}} } ~.
\label{13}
\eeq

The radial wave functions are calculated by integrating the HFB
equations
outwards starting from the initial conditions (12), and inwards
starting from (10) or (11) depending on the value of E.
The solutions are propagated by a modified Numerov method towards the
matching point, where the continuity conditions for the wave functions
and their derivatives are imposed. These conditions determine
the coefficients $D_1$, $D_2$, $D_3$ and the phase shift $\delta$
for a quasiparticle state in the continuum;
in the case of a discrete quasiparticle state the continuity conditions
and the normalisation condition determine the coefficients $A$, $B$,
$D_2$, $D_3$
and the energy E. 

The difficulty of an exact continuum calculation, i.e., with asymptotic
solutions given by Eq.(11), is to identify the energy regions
where
the localisation of the wave functions changes quickly
with the quasiparticle 
energy. These are the regions of quasiparticle resonant states.

In HFB the quasi-particle resonant states are of two types. 
A first type corresponds to the single-particle resonances of the mean
field. 
The low-lying resonances of the mean field located close to the particle 
threshold are very important in the treatment of pairing correlations 
of weakly bound nuclei because they become strongly populated by pairing
correlations.

A second kind of resonant states is specific to the HFB method and
corresponds to the bound 
single-particle states which in the absence of pairing correlations
have an energy $\epsilon < 2\lambda$ . In the presence of the pairing
field these bound states are coupled with the continuum single-particle
states and therefore they acquire a width. The positions and the widths of
these HFB 
resonances are
related to the total phase shift, calculated from the 
matching conditions, as \cite{bulgac}:
\beq
\delta(E) \simeq \delta_0(E) + arctg { \Gamma \over {2(E_R -E)} }~,
\label{62}
\eeq
where $E_R$ and $\Gamma$ are the energy and the width of the resonant 
quasiparticle state. The function $\delta_0(E)$ is the phase shift
of the upper component of the HFB wave function in the HF limit, i.e.,
$\tilde{h}=0$. In this limit one has:
\beq
h \Phi_1 ^0 =(E+ \lambda) \Phi_1 ^0 ~.
\label{60}
\eeq
If there is no single-particle resonance close to
the the energy $E+ \lambda$ in the HF limit, then the HF phase shift
$\delta_0$ has 
a slow variation in the quasiparticle energy region.
In this case the derivative of the total phase shift has a Breit-Wigner 
form, which can be used for estimating the position and the width of
the quasiparticle resonance.

Thus, in the first step of the calculations we study for  
each $(l,j)$ channel the behaviour of the phase shift and we
estimate the energies (widths) of the resonant states from the energies
where the derivative of the phase shift is maximum (half of its maximum).
Then, we choose for the integration in the energy region of a resonant state
an energy grid with a small
step.  In the calculations
presented below the energy step in the region of a resonance is  
$\Gamma/10$ and the energy cut-off is chosen to be minus the
depth of the mean field.

\subsection{ Quasiparticle resonances in a schematic model}

In order to illustrate how one can identify the quasiparticle resonances
in HFB calculations, we take here a simple model \cite{belyaev,bulgac}.
Let us assume that the mean field is given by a square well
potential of depth $V_0$ and radius $a$. The pairing field is
taken also as a constant inside the same radius $a$ and zero outside.
In addition, we suppose that the chemical potential $\lambda$ is given. 
For such a system the radial HFB
equations inside the potential well, i.e., for 	$r\le a$, are:
\begin{eqnarray}
 (\frac{1}{r}\frac{d^2}{dr^2}r - \frac{l(l+1)}{r^2}+\alpha^2) u_{lj}
-\gamma^2 v_{lj}& = & 0~,
\nonumber \\
 ( \frac {1}{r}
\frac {d^2}{dr^2} r
- \frac{l(l+1)}{r^2}+\beta^2) v_{lj}
-\gamma^2 u_{lj} & = & 0~.
\label{eq10}
\end{eqnarray}
where $\alpha^2=\frac{2m}{\hbar^2}(\lambda+E+U_0)$,
$\beta^2=\frac{2m}{\hbar^2}(\lambda-E+U_0)$,
$\gamma^2=\frac{2m}{\hbar^2}\Delta$ and $U_0=-(V_0+V_{so}\vec l. \vec s)$.

The solutions of Eqs.(\ref{eq10}) for any value of the quasiparticle energy
are:
\begin{eqnarray}
u_{lj} & = & A_{+}j_l(k_+r) + A_{-}j_l(k_{-}r)~, \nonumber \\
v_{lj} & = & A_{+}g_{+}j_l(k_+r) + A_{-}g_{-}j_l(k_{-}r)~,
\label{eq11}
\end{eqnarray}
where $j_l$ are spherical Bessel functions,
$k_{\pm}=\frac{2m}{\hbar^2}(U_0+\lambda \pm
(E^2-\Delta^2)^{1/2})$ and
$g_{\pm}=(E \pm (E^2-\Delta^2)^{1/2})/\Delta$.

Outside the potential well 
the HFB equations are decoupled. In this case the type of solutions
depends on the quasiparticle energy. They have the forms given
by Eqs.(10), (11).

 In order to simulate the potential corresponding to a heavy nucleus
 close to the drip line, we take for the model parameters the
 following values: $V_0=45.35$ MeV, $V_{so}=0.5$ MeV, $a=5.2$ fm,
 $\Delta=1$ MeV and $\lambda=-2.0$ MeV.

 Here, we discuss only the quasiparticle resonant solutions induced by 
 the bound single-particle states which are
 specific to the HFB approximation. As a typical example we take  
 the case of $p_{1/2}$ states. In the HF limit, i.e., $\Delta$=0, there
 are two bound states at energies $\epsilon_1 =-32.873$ MeV and
 $\epsilon_2 =-10.698$ MeV. When the pairing field is switched on
 these states become quasiparticle resonant states at  energies
 $E_1=30.889$ MeV and $E_2=8.735$ MeV with corresponding widths
$\Gamma_1=
 0.40 $ keV and $\Gamma_2=24.38$ keV. These values are obtained by
 solving the HFB equations in the complex energy plane with outgoing
 wave boundary conditions. On the real energy axis one should find these 
 two resonances from the phase shift behaviour.  
 In Fig. 1 we show the phase shift (top) and its derivative (bottom) in
 the energy region of the second resonant state. One can see that
 the derivative of the phase shift is maximum at the resonance energy,
 and it drops to half of its maximum value when the energy increases by
 about 25 keV. This shows that the behaviour of the phase shift as a
 function of the real energy E gives accurate information on the
 positions and widths of the quasiparticle resonances.
 From Fig. 1 one can also see that the total phase shift does
 not cross $\pi /2$ at the resonance energy. As discussed
 above, the value of the phase shift associated with the resonance energy
 is actually $\delta_r=\pi/2 +\delta_0$. In this case $\delta_0=1.59$,
 so that the resonance appears when the total phase shift crosses a value
 close
 to $\pi$ and not to $\pi/2$. Thus, in order to identify the resonances
 one can calculate the derivative of the total phase shift and
 search for the local maxima, or calculate the HF phase shift $\delta_0$
 and search for the energies associated to $\delta_r=\pi/2 +\delta_0$.
 For the $2p_{1/2}$ state analysed here, the two procedures give exactly
 the same position of the resonance, but this is not generally the case
 even for a square well potential \cite{bianchini}. In the present
 calculations we localise the resonances by using the derivative
 procedure.
   
\section{Results for Ni isotopes}

In this Section we apply the continuum HFB method to the 
case of Ni isotopes, which have been investigated extensively
both in non-relativistic \cite{tehe} and relativistic Hartree-Bogoliubov 
approximation \cite{meng,vinas}.
 
For the Hartree-Fock field we use the Skyrme interaction SIII 
whereas in the pairing channel we choose a density-dependent 
zero-range interaction:
\beq
V=V_0 \left[ 1- \left( { \rho(r) \over {\rho_0} } \right) ^{\gamma} \right]
\delta(\mathbf{r_1}-\mathbf{r_2}) ~,
\label{36}
\eeq
with the following parameters \cite{tehe}: $V_0=-1128.75$MeV,
$\rho_0=0.134 fm^{-3}$ and $\gamma=1$.

Let us first examine the quasiparticle resonant states for the isotope 
$^{84}$Ni. After convergence of the self-consistent procedure
the chemical potential is $\lambda=-1.104$ MeV. In Table 1 we
show the resonant quasiparticle
energies and the widths calculated from the derivatives of the phase
shift. The quasiparticle states $2d_{3/2}$, $1g_{7/2}$ and $1h_{11/2}$ originate
from single-particle resonances while all the others are related to bound
states. 

As already discussed in the case of the schematic model, the
positions of some resonances may appear for  values of the total
phase shifts which are quite far from $\pi/2$. We take here as
an example
the quasiparticle resonance state corresponding to the bound
state $2p_{1/2}$, which was also analysed in the schematic model.
The resonance energy and the width estimated from the derivative
of the phase shift are $E=7.965$ MeV and $\Gamma=338$ keV.
The value of the HF phase shift  is
in this case $\delta_0 = $ 0.656 so that the total phase shift associated
with the resonance should be $\delta_r \simeq \pi /2 +0.656$. The energy
corresponding to this phase shift is $E=$ 7.707 MeV, which
is smaller than the corresponding value extracted from the maximum of the 
derivative of the phase
shift. This shows that in this case the HF phase
shift has a non-negligible variation in the energy 
region of the resonance. However, in practical HFB calculations a
small shift in the actual position of a resonance induced by the
variation of $\delta_0$ is not essential because this information 
is used only to fix an appropriate energy grid for the energy 
integration.

A special behaviour can be noticed for the resonant continuum in
the $s_{1/2}$ channel. As can be seen in Fig. 2, the occupancy 
in this channel increases starting from $-\lambda$ up
to an energy equal to 1.276 MeV. Therefore, in this channel one 
needs to use a very small energy step close to $-\lambda$ in order to 
get a correct description of the pairing correlations.
Finally one should stress the fact that the contribution to the pairing
correlations of this pronounced resonant structure close to the 
quasiparticle continuum threshold is just the manifestation of the 
loosely bound single-particle state $3s_{1/2}$, which in the HFB approach 
is embedded in the continuum. This structure has nothing to do with the
contribution of
the $s_{1/2}$ single-particle background continuum close to zero
energy, which remains very small.

\subsection{ Comparison between continuum and box HFB calculations}

 In this subsection we analyse the sensitivity of the HFB results
 to the continuum treatment in the vicinity of a drip line, by
 comparing the results provided by  
 continuum and box HFB calculations for the chain of
 neutron-rich Ni isotopes. The energy cut-off 
 is the same in both calculations.
For all box calculations
 presented below the box radius is taken equal to 22.5 fermi. 
Recently, some box calculations have been reported for carbon isotopes with
box radii up to 400 fermi \cite{ben2}. If these large box HFB codes could
also be used for heavier nuclei like the Ni isotopes the differences that we
show here between box and exact results near the drip line might be somewhat
reduced. 

 Lets us first discuss the properties directly related to the
 pairing correlations, i.e., pairing correlation energies and
 pairing densities.

The pairing correlation energies are estimated 
by the difference between the total energies
calculated in
HFB and HF approach,
\beq
E_p = E(HF)-E(HFB)~.
\label{38}
\eeq
The results for continuum and box HFB calculations are shown in
Fig. 3 for all Ni isotopes starting from A=74 up to A=88, which is
the last nucleus with positive two-neutron separation energy, as predicted by
the continuum HFB calculations (see below). 
Up to $^{86}$Ni the quantity E(HF) does not depend on the continuous
single-particle spectrum. The isotope $^{88}$Ni
is not bound in HF and therefore the E(HF) used for
estimating the pairing correlation energy is calculated by using
a box, as in box HFB calculations.
From Fig. 3 one can see that the box HFB calculations start
to overestimate the amount of pairing correlations in the proximity
of the drip line. Thus, in box calculations the pairing energy for
$^{84}$Ni is
about twice that of continuum HFB and it is still increasing
for
$^{86}$Ni, where the continuum HFB calculations predict zero
pairing correlation energy. 

These differences are reflected in the pairing densities,
as shown in Fig. 4 for the isotopes $^{84}$Ni and $^{86}$Ni.
One can notice that the box calculations
overestimate the pairing correlations in the surface region, where
the localisation of the resonance wave functions with high $(lj)$ increases. 
 Thus, in the box calculations the resonant states with 
high $(lj)$ located above the Fermi level are more strongly populated
than the corresponding states calculated by using 
continuum HFB calculations. As an
illustration we consider the occupancy of the single-particle resonance
$g_{7/2}$. In $^{84}$Ni this resonance is located at 3.6 MeV and has a width 
of about 25 keV. If we take an energy interval 3.2 MeV$ \le E \le $4 MeV
around the resonance, we find that the total occupancy of the states 
in the box which are within this interval is  about $ 2\% $
higher than the corresponding occupancy in the continuum calculations.
In box calculations the role of a resonant state is usually taken 
by one state with an energy close to the energy of
the resonance, and this state has maximum localisation inside the nucleus. 
Thus, while in box calculations the pairs can virtually
scatter mainly to that state with maximum localisation, in
continuum HFB calculation the pairs can also scatter to the
neighbouring states whose wave functions are less concentrated 
inside the nucleus. As a result the occupancy of a resonance 
in continuum HFB is smaller. This effect, induced by the width of 
resonant states, is missing in box HFB calculations.

Let us consider now the two-neutron separation energies $S_{2n}$:
\beq
S_{2n} = E(Z,N)-E(Z,N-2) ~,
\label{37}
\eeq
which are plotted in Fig. 5. One can see that in both calculations 
the change of the sign of the two-neutron 
separation energies, i.e., the position of the two-neutron drip line, 
is between $^{88}$Ni and $^{90}$Ni, with a faster drop
in the case of continuum HFB.
The values of $S_{2n}$, evaluated within the two HFB calculations are 
in better agreement one with the other than the corresponding 
values of the pairing correlation energies. 
This is because the differences observed in the pairing correlation 
energies are much reduced when one calculates the differences
appearing in $S_{2n}$.
For the same reason one can see that even a HF calculation gives quite
resonable values for the two-neutron separation energies close to the
drip line. The largest differences between HFB and HF calculations 
appear across the doubly magic isotope $^{78}$Ni. In this case the 
pairing energy changes quickly when two neutrons are removed from 
$1g_{9/2}$ or added to $2d_{5/2}$. Because the hole state has larger degeneracy
than the particle state, the pairing correlations are stronger 
in $^{76}$Ni than in $^{80}$Ni. This explains the asymmetry seen 
in the behaviour of $S_{2n}$ across the doubly magic nucleus $^{78}$Ni.
The fact that the value of $S_{2n}$ predicted by HFB for $^{76-78}$Ni 
is close to the data extrapolated from lighter isotopes
indicates that the pairing interaction used in the calculations is 
quite reasonable, at least for the valence shell N=28-50.

Next, we compare the results given by the two HFB calculations for
observables related to mean field properties.
In Fig. 6 the particle density for the isotope
 $^{86}$Ni is shown. One notices that the particle densities are
practically the same except in the region near the box radius.
The fact that the two particle densities are very close up to very large
distances implies that the neutron root mean square radii (rms) calculated 
within the two approaches should be similar. This can be seen in Fig. 
7 for the isotopes $^{80-90}$Ni. 

In Fig. 7 the HF radii are also shown. 
In $^{84}$Ni we can see that the HFB radius is slightly
larger than the HF value, which is the trend usually
expected when the pairing interaction is switched
on. In this case the HFB radius is increased because the
pairing interaction scatters some neutrons from $2d_{5/2}$ to the 
loosely bound state $3s_{1/2}$ which is a state less localised 
inside the nucleus. 
On the other hand, as seen in Fig. 7, the effect of pairing 
correlations on the radius of $^{86}$Ni is opposite. 
Here, the pairing interaction
scatters particles out of $3s_{1/2}$ state which is completely 
occupied in HF. The particles are scattered in the continuum
single-particle states, mainly to single-particle resonances
which have a larger localisation inside the nucleus than the 
$3s_{1/2}$ state.
Thus, in this case the radius is decreased when the pairing correlations
are swiched on. This effect of the pairing interaction on nuclear radii
is sometimes called "anti-halo" \cite{fay,ben}. 
 
\subsection {Comparison between HFB and HF-BCS approximation}

 The HF-BCS approximation is obtained by neglecting in the HFB equations
 the non-diagonal matrix elements of the pairing field. This
 means that  in the HF-BCS limit one neglects the pairing correlations
 induced by the pairs formed in states which are not time-reversed
 partners.

 The extension of BCS equations for taking into account the 
 continuum
 coupling was proposed in Refs.\cite{saliwy,sagili,sacili}. For the
 case of a general pairing interaction the BCS equations read
 \cite{sagili}:
 
\begin{equation}\label{eq:gapr1}
\Delta_i = \sum_{j}V_{i\overline{i}j\overline{j}} u_j v_j +
\sum_\nu
V_{i\overline{i},\nu\epsilon_\nu\overline{\nu\epsilon_\nu}}
\int_{I_\nu} g_{\nu}(\epsilon)
u_\nu(\epsilon) v_\nu(\epsilon) d\epsilon~,
\end{equation}
\begin{equation}\label{eq:gapr}
\Delta_\nu \equiv
\sum_{j}
V_{\nu\epsilon_\nu\overline{\nu\epsilon_\nu},j\overline{j}} u_j v_j +
\sum_{\nu^\prime}
V_{\nu\epsilon_\nu\overline{\nu\epsilon_\nu},
\nu^\prime\epsilon_{\nu^\prime}
\overline{\nu^\prime\epsilon_{\nu^\prime}}}
\int_{I_{\nu^\prime}} g_{\nu^\prime}(\epsilon^\prime)
u_{\nu^\prime}(\epsilon^\prime) v_{\nu^\prime}(\epsilon^\prime)
d\epsilon^\prime~,
\end{equation}
\begin{eqnarray}
N = \sum_i v_i^2 + \sum_\nu \int_{I_\nu} g_{\nu}(\epsilon) v^2_\nu
(\epsilon) d\epsilon~.
\label{eq15}
\end{eqnarray}
Here $\Delta_i$ is the gap for the bound state $i$ and
$\Delta_\nu$ is the averaged gap for the resonant state $\nu$.
The quantity $g_\nu(\epsilon) = \frac {2j_\nu +1}{\pi}
\frac{d\delta_\nu}{d\epsilon}$ is the continuum level density 
and
$\delta_\nu$ is the phase shift of angular momentum $(l_{\nu} j_{\nu})$.
The factor $g_\nu(\epsilon)$ takes into account the variation
of the localisation of scattering states in the energy region of
a resonance ( i.e., the width effect) and becomes a delta function in the
limit of a very narrow width. 
In these equations the interaction matrix elements are  calculated with
the scattering wave functions at resonance energies and normalised
inside the volume where the pairing interaction is active. 
The BCS equations (21-23) are solved iteratively 
together with the HF equations. The corresponding equations are
called below the resonant continuum HF-BCS equations. For more
details see Ref.\cite{sagili}.

In the case of Ni isotopes the effect of the continuum is introduced
through the first three low-lying single-particle
resonances, i.e., $d_{3/2}$, $g_{7/2}$ and $h_{11/2}$. These resonances
form together with the bound states $2d_{5/2}$ and $3s_{1/2}$
the equivalent of the major shell $N=50-82$.
The energy integrals in BCS equations (21-23) are
performed for each resonance in an energy interval defined such that
$\vert \epsilon - \epsilon_\nu \vert \leq 2\Gamma_\nu $, where
$\epsilon_\nu$ is the energy of the resonance and $\Gamma_\nu$ is
its width. In the resonant continuum HF-BCS calculations we use the
same interaction as in HFB approach. 

In Fig. 8 we show the pairing correlation energies  predicted by
the resonant continuum HF-BCS
approximation in comparison with the continuum HFB results.
One can see that the HF-BCS results follow closely the exact HFB
values up to the drip line. This shows that in order to estimate the
pairing
correlations one needs  to include from the whole continuum 
only a few resonant states with their widths properly considered. 

In order to see the effect of the widths of resonant states upon pairing, 
we replace in the resonant continuum HF-BCS
equations the continuum level density by delta functions. 
This means that the resonant state is replaced by a scattering state at the 
resonance energy,
 normalized in a volume 
of radius R. For this radius we take the same value
as in box HFB calculations, i.e. 
R=22.5 fm. As it can be seen from Fig. 9, the pairing correlations
increase
when one neglects the widths of the resonances and the results follow  closely 
those of box HFB
calculations.
Thus, the overestimation of pairing correlations due to the continuum
discretisation is similar in HF-BCS and HFB calculations.

 In Fig. 7 we show also the radii calculated in the resonant continuum
HF-BCS approximation. One notices that the HF-BCS radii are closer to 
the HF values than to the HFB ones.  The same behaviour is found
for the particle densities. This can be seen in Fig. 6 for the
case of $^{86}$Ni, which is the last bound nucleus in
the HF approximation. From Fig. 6 one can see also that the tail 
of the density in resonant continuum HF-BCS calculations is mainly 
given by the particles distributed in the bound states $2d_{5/2}$ 
and $3s_{1/2}$ and not due to the particles scattered to positive energy
states. In HFB calculations a part of the
particles from the bound states $2d_{5/2}$ and $3s_{1/2}$ are scattered to
other states, mainly to resonant states, with wave functions 
concentrated inside the nucleus. Therefore the HFB density has a smaller
tail at large distances. 

As we have already mentioned, in the present resonant continuum
HF-BCS calculations we
neglect all the continuum contribution except for the three
low-lying resonances $d_{3/2}$, $g_{7/2}$ and $h_{11/2}$. 
This model space seems sufficient for a proper evaluation of pairing
correlation energies up to the drip line. The rest of the continuum
 changes mainly the particle distribution.
In order to get a particle density closer to the HFB results
one needs to introduce in the resonant continuum HF-BCS 
calculations aditional relevant pieces from the continuum. 
This work is in progress.

\section{Conclusions}

 In this article we have discussed how one can actually solve the 
 HFB equations with proper boundary conditions for the continuous
 spectrum and we have shown, for the case of  neutron-rich Ni isotopes,
 how different treatments of the continuum can affect the pairing
 correlations.
 It was found that in the vicinity of the drip line pairing
correlations are overestimated
by the continuum discretisation done in box HFB calculations.
On the other hand, we have shown that the particle densities and the radii
are rather insensitive to the
 way in which the continuum
is treated in HFB calculations. This means that the quantities that are
mainly related to the mean field properties do not practically depend on the
different treatments of continuum. We have also shown that the position of
the two-neutron drip line is not affected by the way in which continuum is
treated. This is due to the fact that the differences observed for the 
pairing correlations energies in the two HFB calculations are diminished
when the two-neutron separation energies are calculated. Moreover,
the two-neutron separation energies predicted by HF are not very different 
from the HFB results. This shows that these quantities
 are not indicated for testing the pairing
correlations close to the drip line.

 We have also analysed how the exact HFB solutions compare to
 the resonant continuum HF-BCS approximation \cite{saliwy,sagili}. 
 It was shown that the resonant HF-BCS calculations
 which include only the first three low-lying resonances
 provide a very good description of pairing correlation energies up
 to the drip line. On the other hand, in the vicinity of the drip line
 the radii predicted by the resonant continuum HF-BCS
 calculations are larger than the HFB radii and closer to the
 HF results. 
 This shows that one should add to the first three low-lying
 resonances additional contributions of the continuum in order to evaluate
 better the particle densities for nuclei close to the drip line.

\begin{acknowledgments}
We thank J. Dobaczewski for providing us the code which
solves the HFB equations with box boundary conditions.
One of us (N.S.) would like to thank the Insitute de Physique
Nucl\'eaire - Orsay for its hospitality.
This work was done in the framework of IN2P3-IPNE
Collaboration.

\end{acknowledgments}

\newpage

\begin{table}[h]
\label{table1}
\caption{Hartree-Fock single-particle energies $\epsilon$, HFB
quasiparticle resonance energies (E) and widths ($\Gamma$) in the
nucleus $^{84}$Ni, for the various $(lj)$ states involved.}

\vspace{0.2cm}
\begin{center}
\footnotesize
\begin{tabular}{||c|c|c|c|c||}
\hline
l & j & $\epsilon$ (MeV) & E (MeV) & $\Gamma (keV)$ \\
\hline
0 & 1/2 & -0.731 & 1.276 & \\
\hline
 & & -22.530 & 20.878 & 98  \\ 
 \hline
  & & -45.010 & 43.3917 & 0.3  \\ 
  \hline
  1 & 1/2 & -9.540 & 7.965 & 338  \\ 
  \hline
   & & -34.709 & 33.444 & 102  \\ 
   \hline
    1 & 3/2 & -11.194 & 9.712 & 576  \\ 
    \hline
     & & -36.364 & 34.976 & 76  \\ 
     \hline
      2 & 3/2 & 0.475 & 2.317 & 816  \\ 
      \hline
        & & -23.055 & 22.028 & 58  \\ 
	\hline
	 2 & 5/2 & -1.467 & 1.845 & 44  \\ 
	 \hline
	   & & -26.961 & 25.628 & 3  \\ 
	   \hline
	    3 & 5/2 & -10.586 & 8.863 & 944  \\ 
	    \hline
	    3 & 7/2 & -17.023 & 15.857 & 882  \\ 
	    \hline
	     4 & 7/2 & 1.604 & 3.598 & 24 \\ 
	     \hline
	      4 & 9/2 & -6.837 & 5.674 & 3  \\ 
	      \hline
	      5 & 11/2 & 3.295 & 5.380 & 52  \\ 
	      \hline
	      \end{tabular}
	      \end{center}
	      \end{table}

\newpage

\begin{figure}
\caption{Phase shift (top) and its derivative (bottom) in the $p_{1/2}$ channel
for a square well model.}
\label{1a}
\end{figure}

\begin{figure}
\caption{Occupation probability profile in the 
 $s_{1/2}$ channel for $^{84}$Ni.}
\label{2a}
\end{figure}

\begin{figure}
\caption{Pairing correlation energies for Ni isotopes calculated in HFB
approximation. }
\label{3a}
\end{figure}

\begin{figure}
\caption{Neutron pairing densities in in HFB calculations
in $^{84}$Ni (a) and $^{86}$Ni (b).}
\label{4a}
\end{figure}

\begin{figure}
\caption{Two-neutron separation energies in HFB, HF-BCS and HF 
approximations. For $^{76}$Ni and $^{78}$Ni the corresponding values
extrapolated from experimental data [18] are also shown.}
\label{5a}
\end{figure}

\begin{figure}
\caption{Neutron particle densities in HFB, resonant continuum
 HF-BCS and HF approximations for $^{86}$Ni. The density represented by the 
 dotted
 line (HF-BCS bound)
 is calculated by including  only
 the contribution of bound states.}
\label{6a}
\end{figure}

\begin{figure}
\caption{Neutron rms for Ni isotopes in HFB, resonant continuum
 HF-BCS and HF approximations.}
\label{7a}
\end{figure}

\begin{figure}
\caption{Pairing correlation energies calculated in resonant
continuum HF-BCS approximation compared to continuum HFB.}
\label{8a}
\end{figure}

\begin{figure}
\caption{Pairing correlation energies calculated in the resonant
continuum HF-BCS approximation by neglecting the widths effect  
compared to box HFB results.}
\label{9a}
\end{figure}

\newpage

\includegraphics{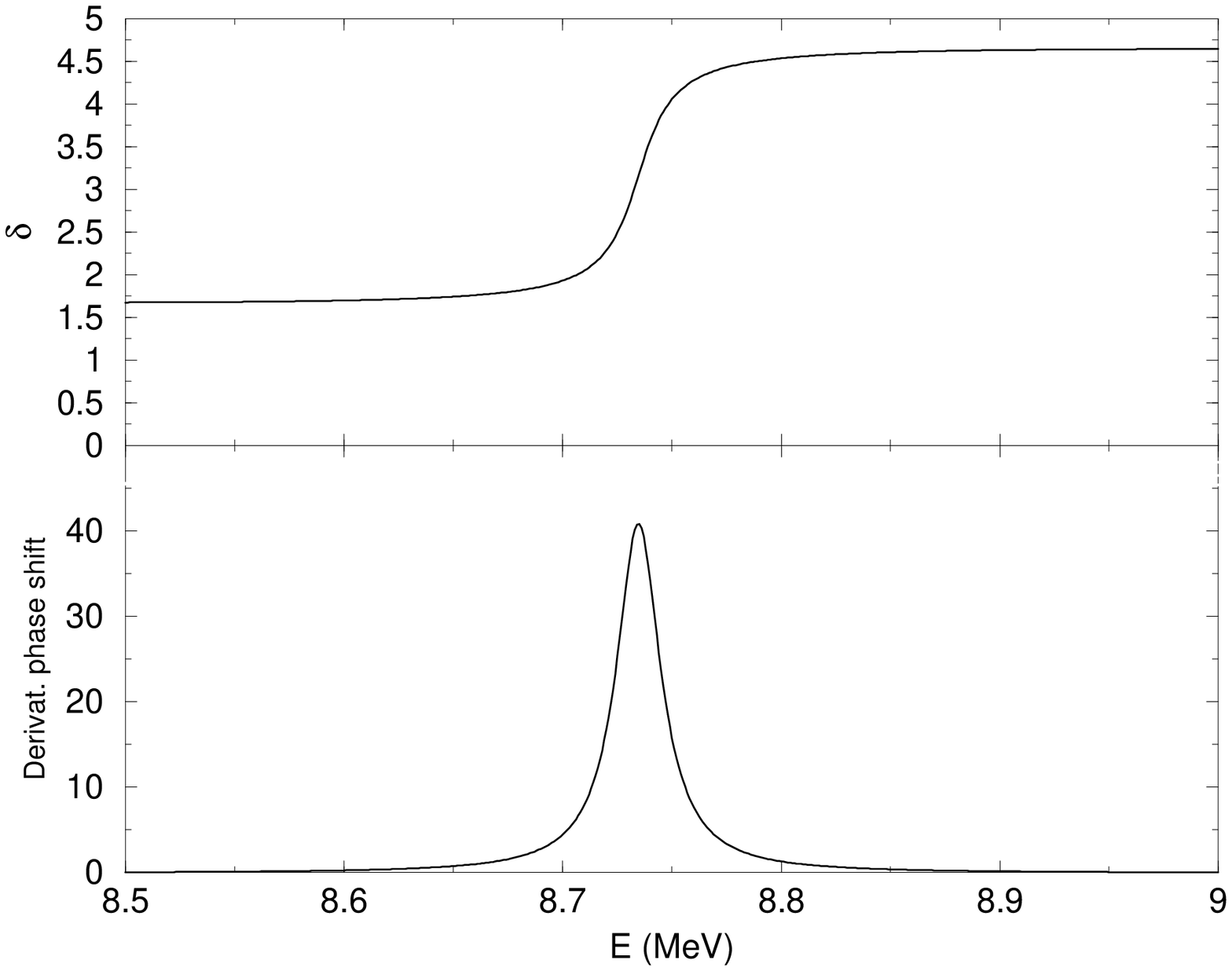}
\mbox{}\\

\newpage

\includegraphics{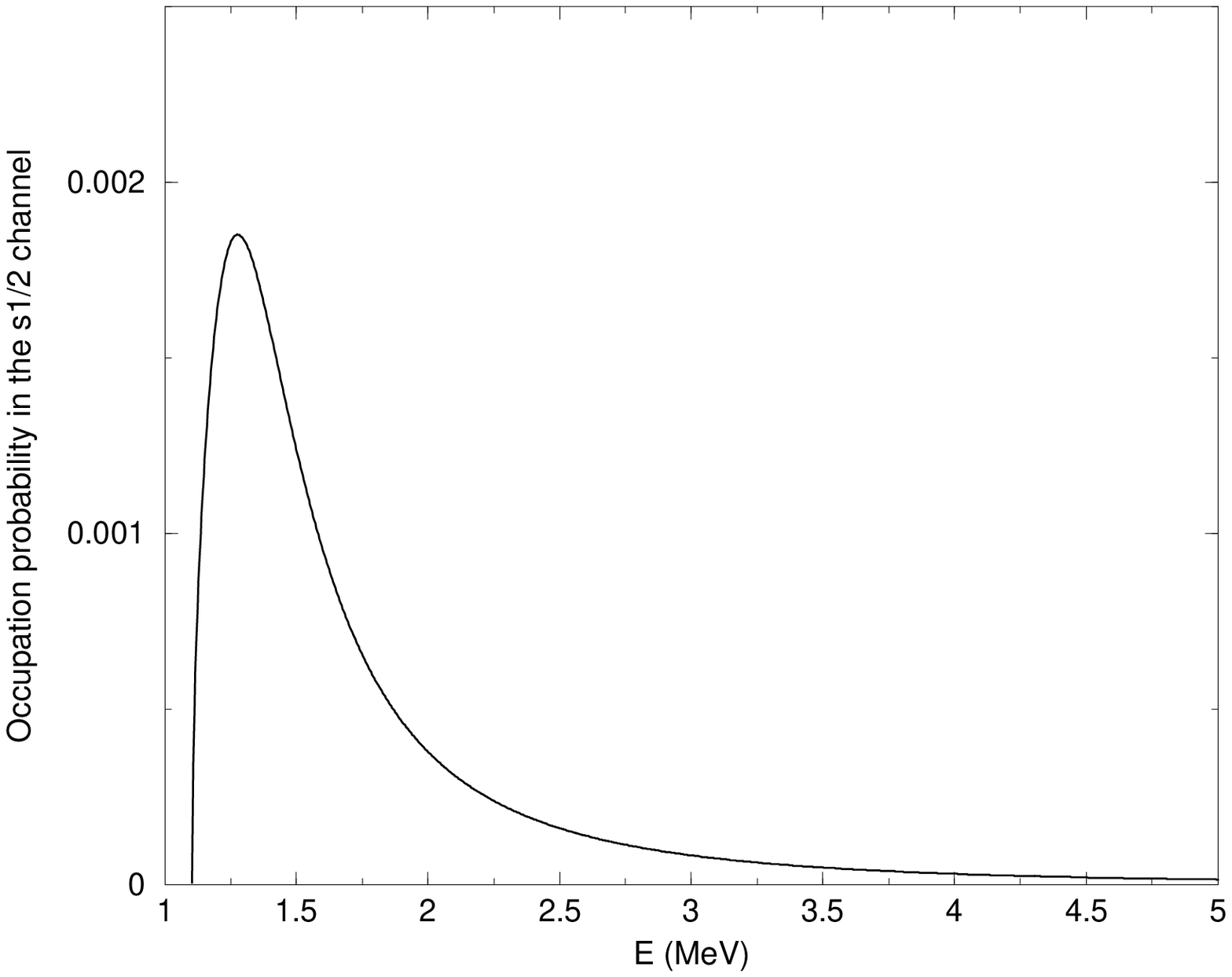}
\mbox{}\\

\newpage

\includegraphics{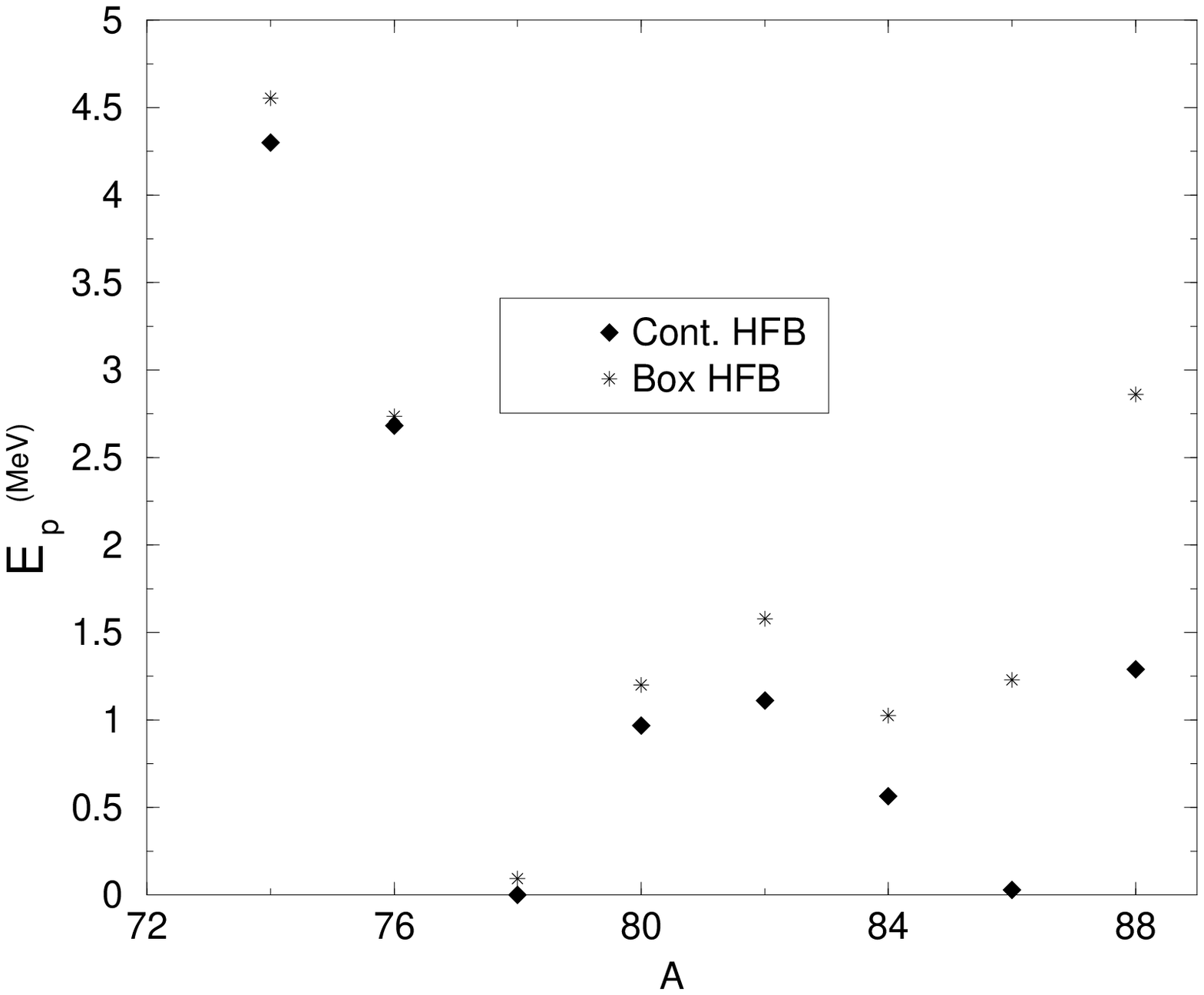}
\mbox{}\\

\newpage

\includegraphics{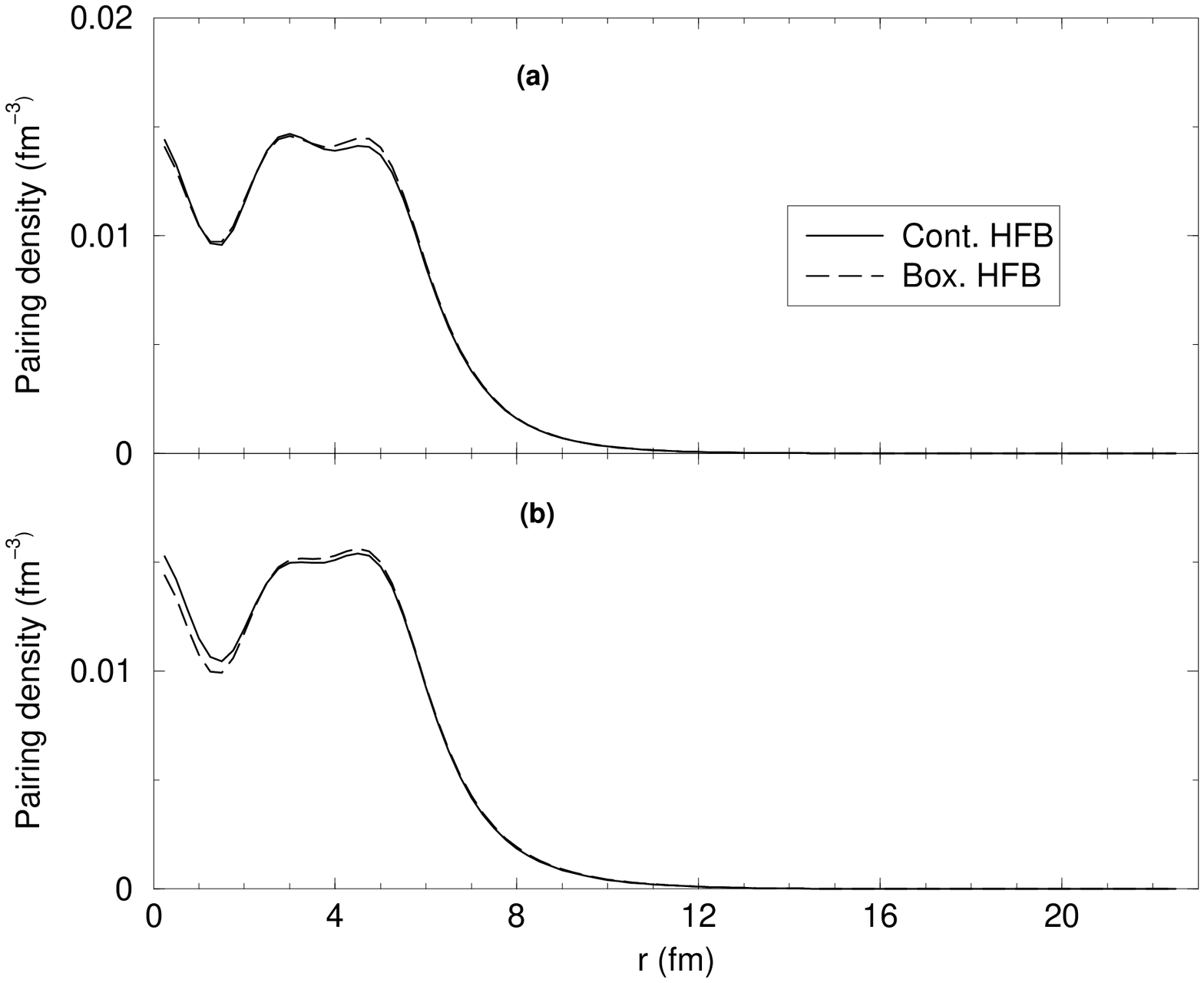}
\mbox{}\\

\newpage

\includegraphics{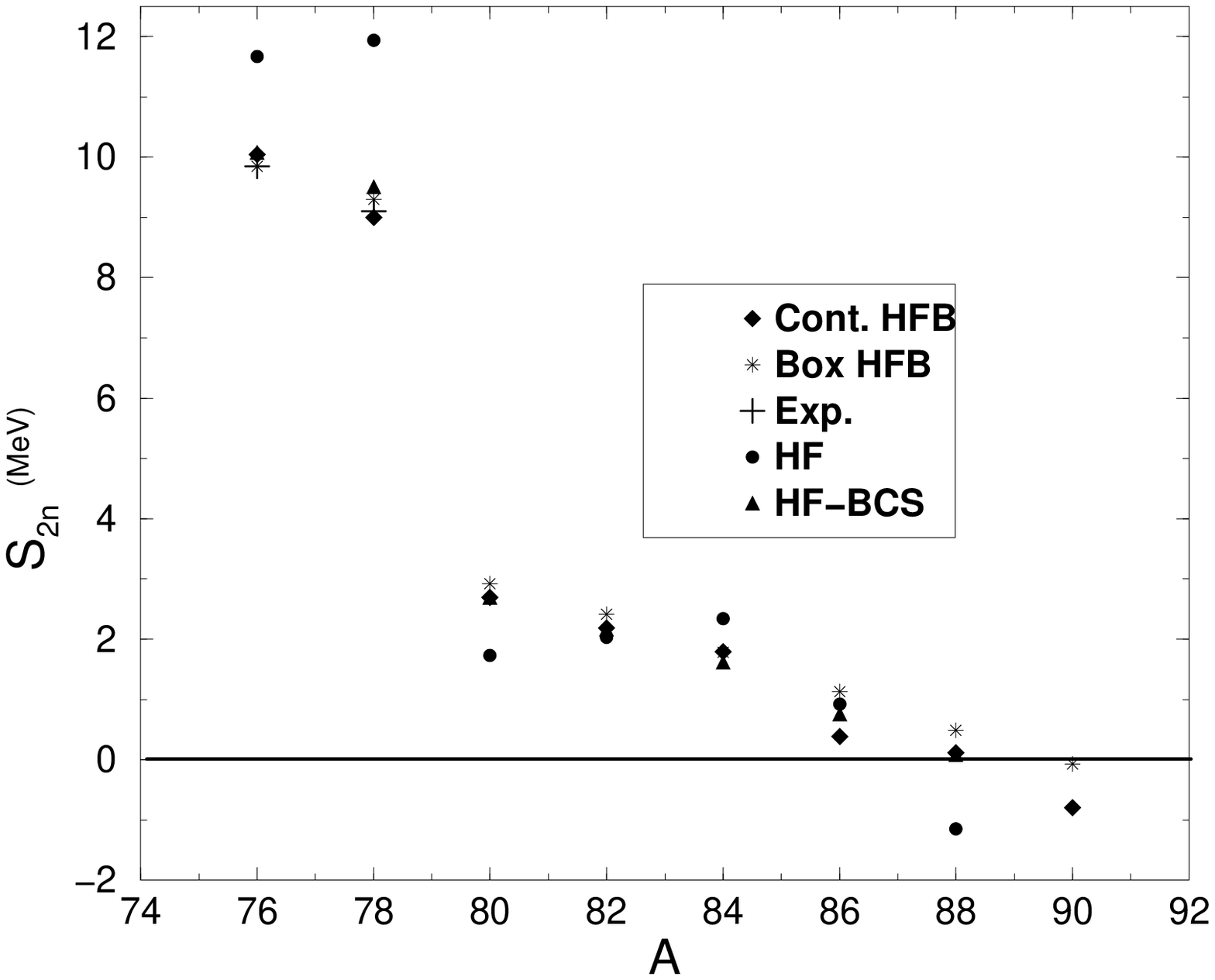}
\mbox{}\\

\newpage

\includegraphics{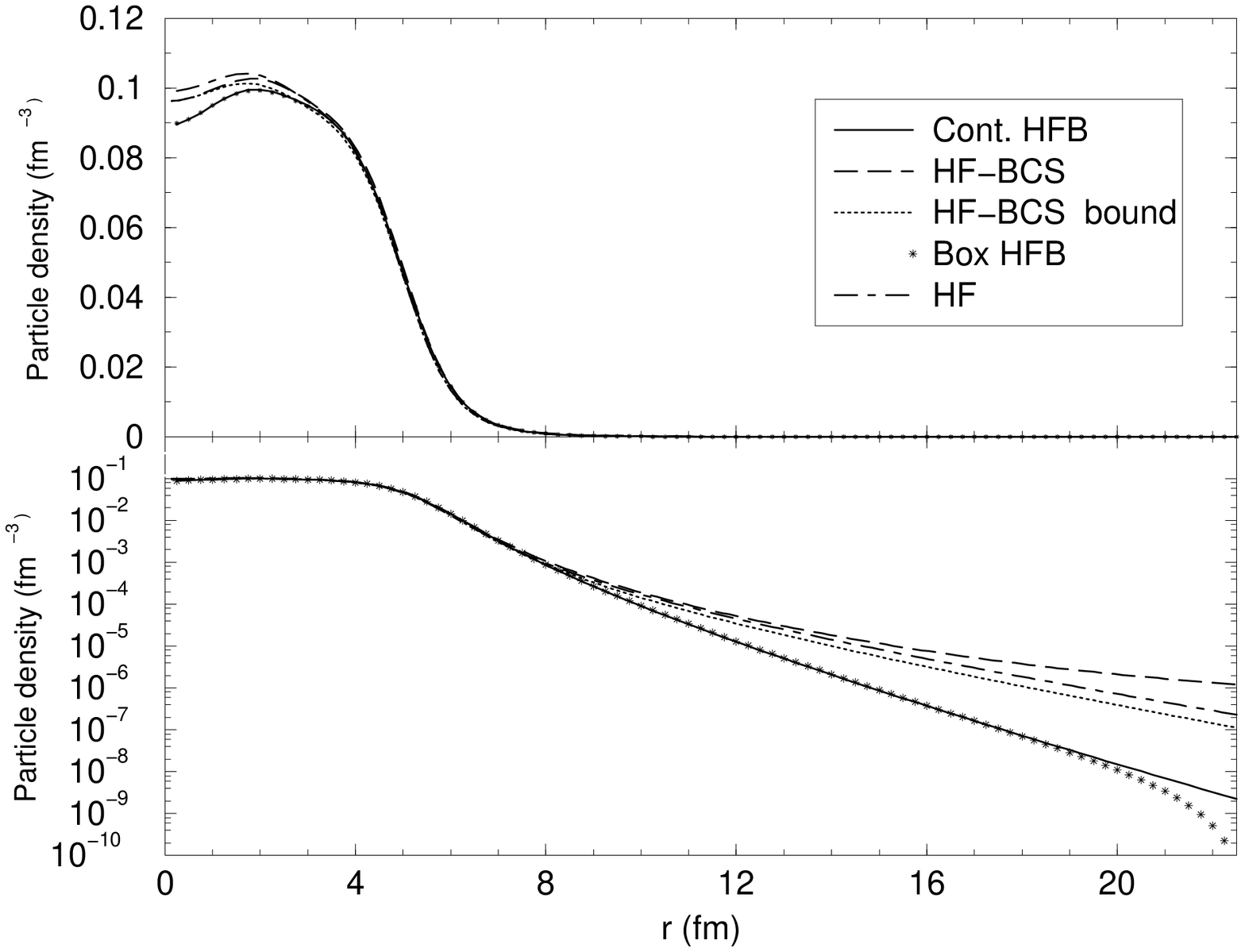}
\mbox{}\\

\newpage

\includegraphics{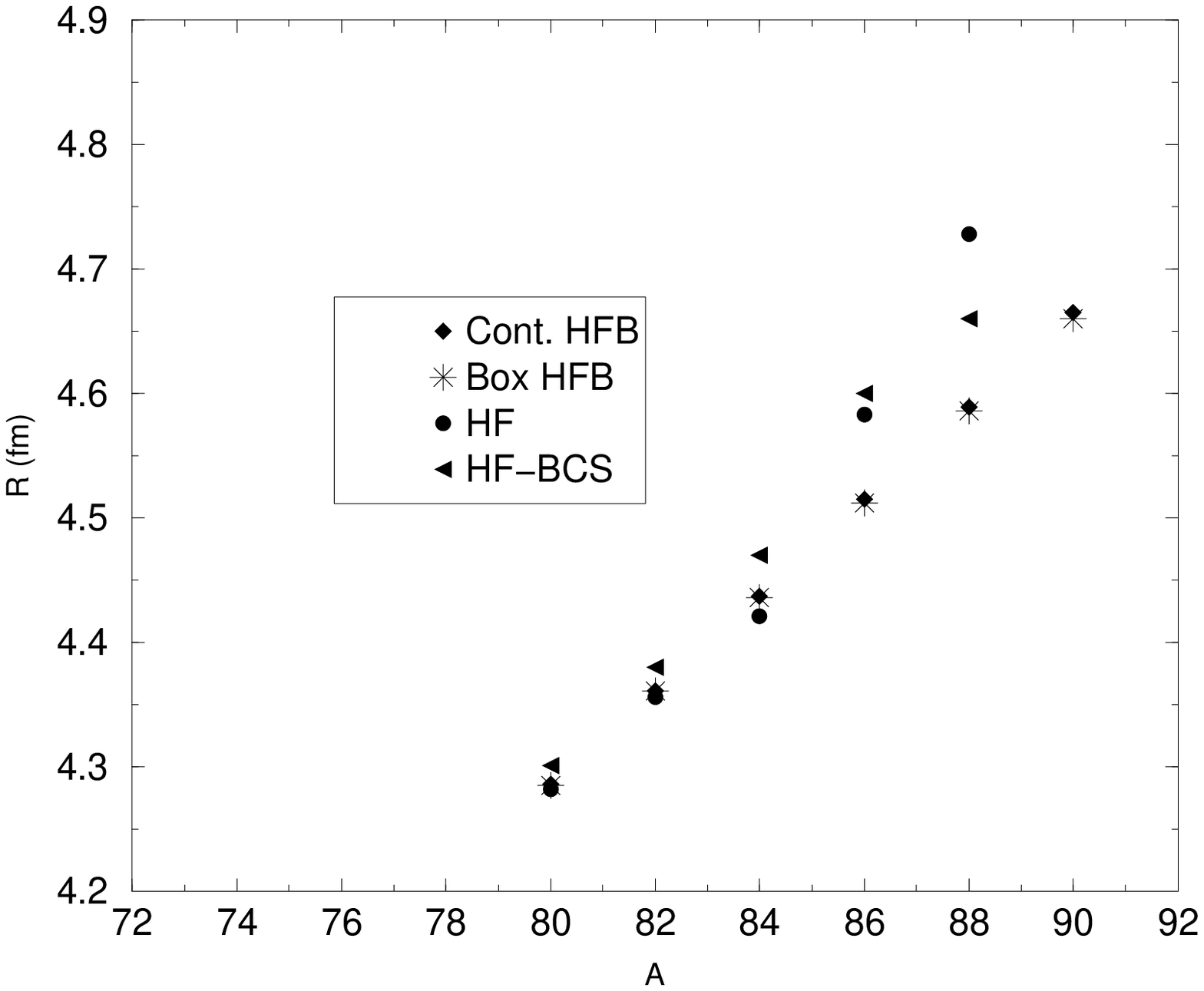}
\mbox{}\\

\newpage

\includegraphics{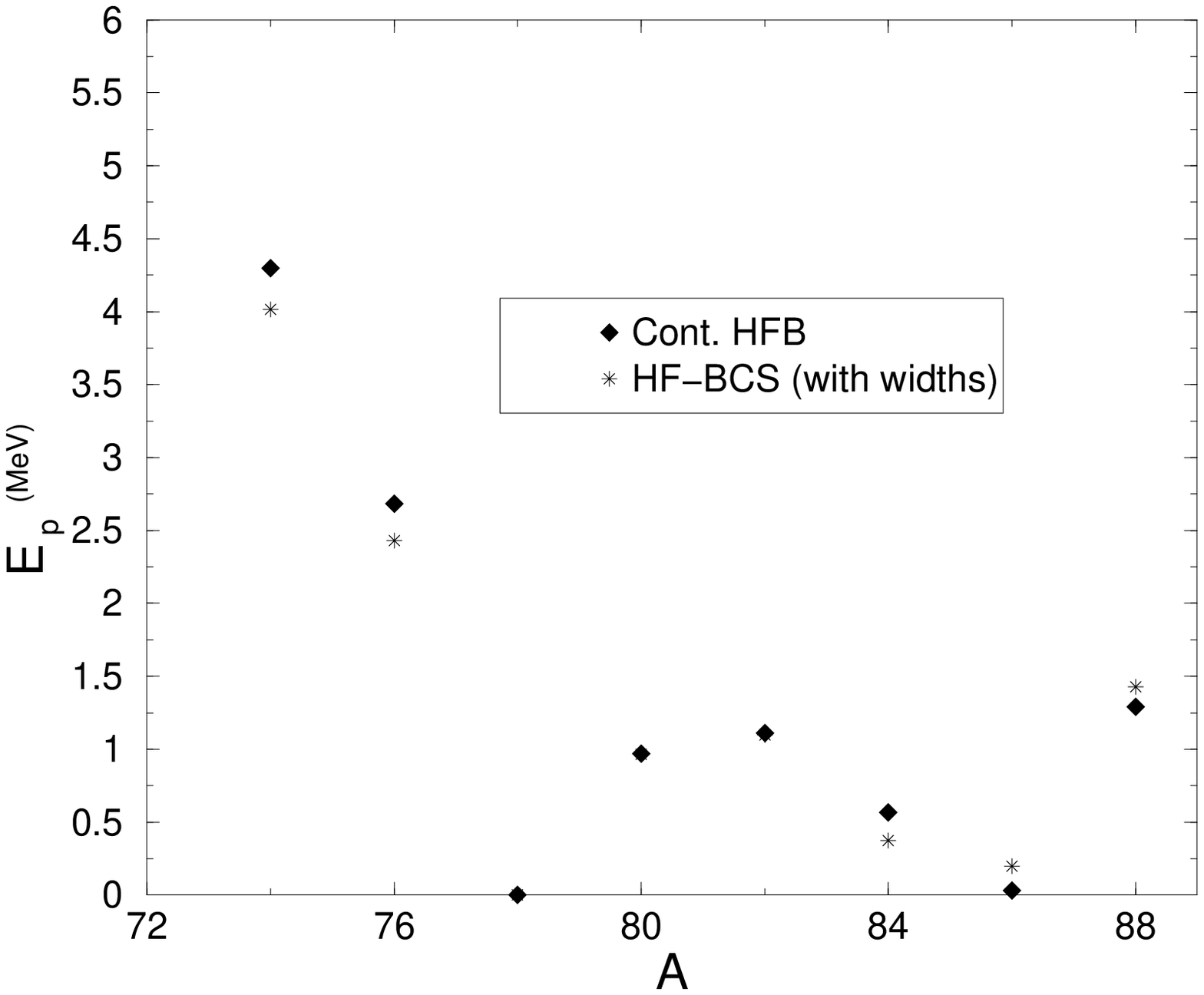}
\mbox{}\\

\newpage

\includegraphics{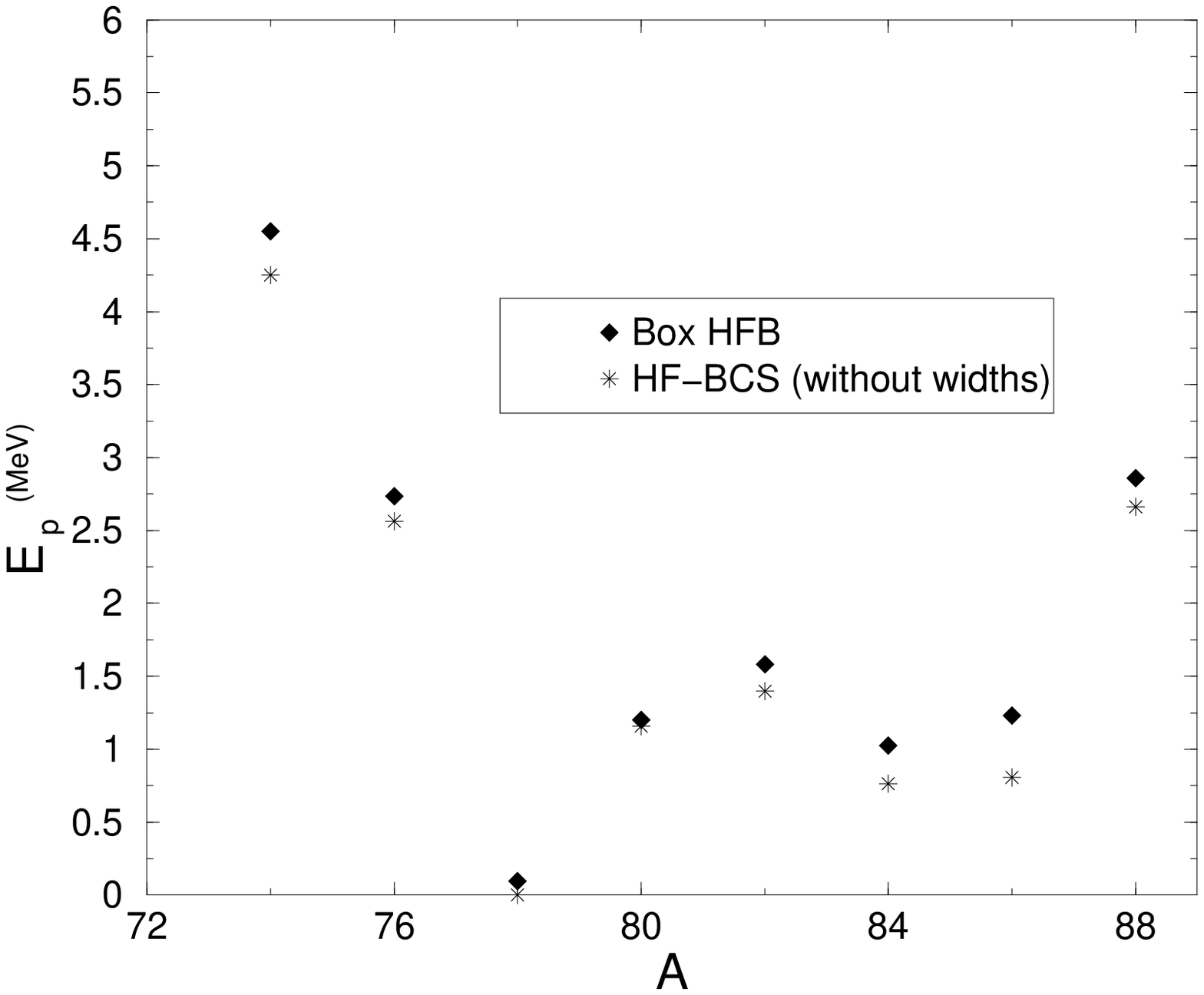}
\mbox{}\\

\end{document}